\documentclass[9pt,twocolumn,twoside]{pnas-new}

\templatetype{pnasresearcharticle} 

\title{Generalized Network Dismantling}

\author[a,1]{Xiao-Long Ren}
\author[b,1]{Niels Gleinig} 
\author[a]{Dirk Helbing}
\author[a,2]{Nino Antulov-Fantulin}

\affil[a]{Computational Social Science, ETH Z\"urich, Clausiusstra{\ss}e 50, 8092 Z\"urich, Switzerland}
\affil[b]{Department of Computer Science, ETH Z\"urich, Switzerland}

\leadauthor{Ren} 

\significancestatement{
The functioning of many socio-technical systems depends on the ability of its subcomponents or nodes to communicate or interact via its connections.
By removing or deactivating a specific set of nodes, 
a network structure can be dismantled into isolated subcomponents, thereby disrupting the malfunctioning of a system or containing the spread of misinformation or an epidemic.
We propose a generalized network dismantling framework, which can take realistic removal costs into account such as the node price, protection level or removal energy.
We discuss applications of cost-efficient dismantling strategies to real-world problems such as epidemic spread or criminal and corruption networks.
}

\authordeclaration{No conflict of interest.}
\equalauthors{\textsuperscript{1}X.-L.R. and N. G. contributed equally to this work.}
\correspondingauthor{\textsuperscript{2}To whom correspondence should be addressed. E-mail: anino@ethz.ch}


\begin{abstract}
Finding an optimal subset of nodes in a network that is able to disrupt the functioning of a corrupt or criminal organization or contain an epidemic or the spread of misinformation is still one of the open problems in modern network science. 
In this paper, we introduce the generalized network dismantling problem, which aims at finding a minimum set of nodes that, when removed from a network, results in the fragmentation of a network into subcritical network components at minimum cost.
Contrary to previous formulations, we allow the costs for node removal to take arbitrary non-negative real value.
For unit costs, our formulation becomes equivalent to the standard network dismantling problem. Our non-unit cost generalization allows one to consider topological cost functions related to node centrality or non-topological features such as the price or protection level of a node. 
In order to solve this optimization problem, we propose a method, which is based on the spectral properties of a novel node-weighted Laplacian operator.  
The proposed method is applicable to large-scale networks with millions of nodes. It outperforms current state-of-the-art methods and opens new directions for understanding the vulnerability and robustness of complex systems. 
\end{abstract}

\begin{document}

\verticaladjustment{-2pt}

\maketitle
\thispagestyle{firststyle}
\ifthenelse{\boolean{shortarticle}}{\ifthenelse{\boolean{singlecolumn}}{\abscontentformatted}{\abscontent}}{}

\dropcap{I}n a hyper-connected world, systemic instability, based on cascading effects, can seriously undermine the functionality
of a network\cite{Helbing2013}. 
The quick global spread of rumors and fake news may be seen as recent examples \cite{DelVicario2016, Waldrop2017, Lazer2018}, while the spread of epidemics \cite{Vespignani2011, Brockmann2013, AntulovFantulin2015} or failure propagation \cite{CascadeFailure, CascadeFailure3, CascadeFailure4} is a problem that has been around much longer. 
Furthermore, it is known that the network structure, for example the exponent characterizing 
scale-free networks, is of particular importance for the controllability of cascading effects \cite{Dorogovtsev2008}.
For certain scaling exponents of scale-free network, the variance or mean value of relevant quantities may not be well-defined, which means that unpredictable or uncontrollable
behavior may result. It may then be impossible to contain epidemic spreading processes. Similar circumstances may make
it impossible to contain the spread of computer viruses or misinformation---a problem that is not only relevant for the quick increase
of cyberthreats, but may also undermine the functionality of markets, societal or political institutions. 
At the same time, the removal or deactivation of even a small set of collective influential nodes can dismantle the network into isolated subcomponents and thus disrupt the malfunctioning of a system.
For example, scale-free networks~\cite{BA, Dorogovtsev2000} are more robust to random removals than Erdős–Rényi network~\cite{ER,Gilbert1959}, but at the same time more vulnerable to targeted attacks \cite{Molloy1995, Albert2000, PhysRevLett.85.4626, Cohen2001, Tanizawa2005, SchneiderPNAS, Gallos2006AttackStrategies}.

However, finding a minimum set of nodes, the removal of which is able to dismantle a network \cite{Braunstein2016, Morone2015Nature} into isolated subcomponents of specific small size belongs to class of hard computational problems, called non-deterministic polynomial hard (NP-hard) problems. Essentially, this implies that there currently exists no efficient algorithm that can find the best dismantling solution for large-scale structures. However, this does not exclude finding approximate dismantling solutions. For example,
novel approximations \cite{Morone2015Nature, Kovcs2015, Braunstein2016, Zdeborova2016, BPAttack, Morone2016, Tian2017NatComm, Zhou2013FVS}, based on spin-glass and optimal percolation theory, have been proposed. But still, all these methods make the implicit assumption that the cost of removing nodes is constant. Only recently, the effect of costs depending on the node degree was studied \cite{ShlomoDegreeAttack, NcutAttackArxiv}, but restricted to random network structures \cite{ShlomoDegreeAttack} or edge-based strategies \cite{NcutAttackArxiv} that take the degree-based cost into account. 

This paper addresses the question of how to select the set of nodes in a network, that, when removed or (de)activated, can stop the spread of (dis)information, or an epidemic or disrupt the functioning of a malicious systems by fragmenting it into small components with minimal cost.
In the generalized network dismantling problem, the cost of removing a node can be an arbitrary non-negative real number, which can, for example, be specified as a function of node centrality properties \cite{Lu2016} such as degree, PageRank, betweenness or other variables unrelated to network topology. Examples of variables that may determine the cost of removing a node include: the monetary price for buying or controlling a node, the protection level or the energy effort.
In this paper, we reformulate the Laplacian spectral partitioning~\cite{fiedler73, Pothen1990}, which dates back to 1973 and since then was primarily used for edge removal strategies on networks.
We then construct a special node--weighted Laplacian operator, which serves to determine the upper bound of the cost of the generalized network dismantling problem. Furthermore, we propose an elegant and efficient approximation algorithm for the problem, which is applicable to large-scale networks. Finally, a fine-tunning mechanism is introduced by mapping the spectral solution to the weighted vertex cover problem \cite{BARYEHUDA1981} from graph theory. 
Understanding the 
key relationships between dismantling solutions and their cost(s) enables one
to increase the level of robustness of real-world systems.

\begin{figure*}
\centering
\includegraphics[width=\linewidth]{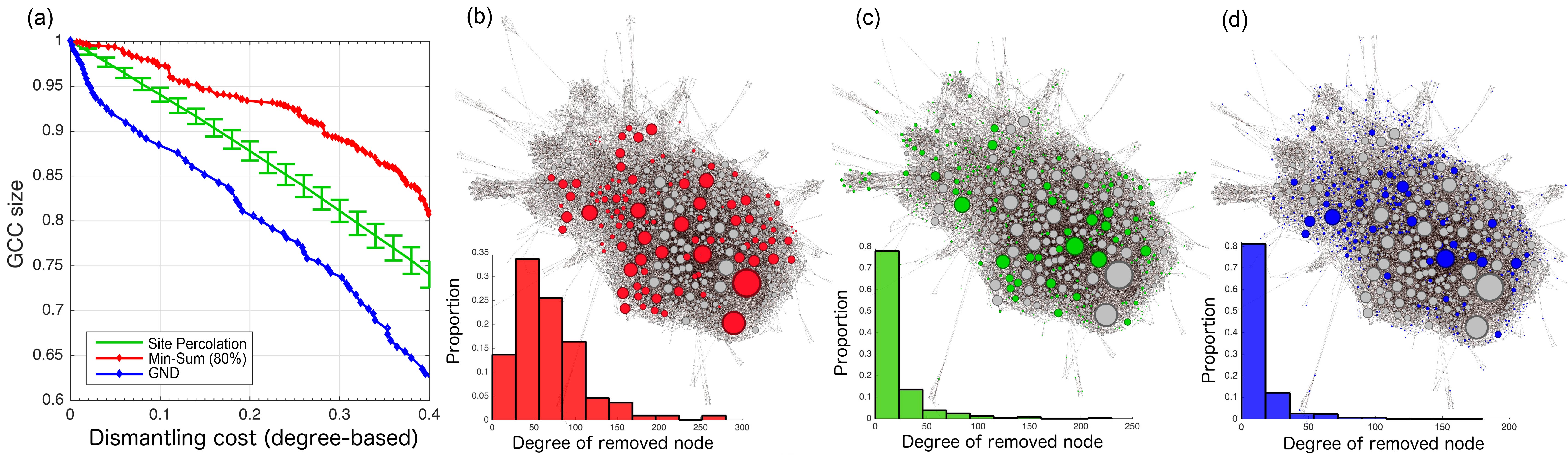}
\caption{\textbf{(a)} Network dismantling measured by the size of the giant connected component (GCC) with respect to the degree-based definition of the cost for three different strategies: state-of-the-art Min-Sum \cite{Braunstein2016}, random removal \cite{Callaway2000} (site percolation), our GND strategy. 
Dismantling represents the controlled process of suppressing the spread of misinformation, computer viruses or other harmful contagion effects on a online social network (Petster-hamster \cite{KONECT}).
The Min-Sum algorithm was set to dismantle the network up to a fixed target size of 80 percent of the network GCC size. The cost of removing a node is proportional to the current degree of a node and the cost of the dismantling is measured with the fraction of removed edges adjacent to the removed nodes. 
We observe that, for the same dismantling cost 0.4, the Min-Sum algorithm produces results (red color) which are 5 $\%$ worse than the naive random removal (green color) of nodes in a network. However, for the same cost the proposed GND strategy (blue color) fragments the network up to 62 $\%$ GCC size, which is 18 $\%$ better than the Min-Sum strategy.
\textbf{(b,c,d)} Online social network (Petster-hamster \cite{KONECT}) displaying the set of removed nodes according to the Min-Sum (red color), random removal (green color), and our GND strategy (blue color). Although the Min-Sum strategy removes a small percentage of nodes, the cost is rather high as it targets high degree nodes, which is visible from the histogram of the removed nodes. In contrast, the GND strategy avoids the expensive removal of hubs in this scenario and produce a much better fragmentation. }
\label{fig:motivation}
\end{figure*}

\section*{Main Contributions}
The main contributions of this paper are listed below:
\begin{enumerate}[label=(\roman*)]
\item We introduce a generalized network dismantling problem, which seeks to find a set of nodes that, when removed from a network, results in a network fragmentation into components of subcritical size at minimum cost. Contrary to the previous formulations \cite{Morone2015Nature, Kovcs2015, Braunstein2016, Zdeborova2016, BPAttack, Morone2016, Tian2017NatComm, Zhou2013FVS}, assuming identical costs for the removal of each node, we allow for costs that have arbitrary non-negative real values. 
\item We formulate a novel node-weighted graph-cut objective function, which determines the upper bound for the generalized network dismantling cost. We find the analytical solution for the relaxed objective function, which is related to the spectral properties of the node-weighted Laplacian matrix. 
\item To dismantle large-scale networks, we propose an efficient spectral approximation by constructing a Power Laplacian operator, which has complexity $O( n \cdot \log^{2+\epsilon}(n))$. Furthermore, we provide analytical bounds and convergence proofs for the spectral approximation. Finally, we propose a fine-tuning mechanism by mapping the problem to the weighted vertex cover problem.  
\item We show that, on real networks, our approach outperforms current state-of-the-art methods \cite{Morone2015Nature, Braunstein2016, Zdeborova2016, Morone2016, Tian2017NatComm} for non-unit costs.
In unit cost scenario, our approach performs either better or
comparably to other state-of-the-art methods. 
\end{enumerate}

\section*{Generalized network dismantling problem}
Let us define a network $G(V,E)$ as the set of nodes, $V$, which are connected via a set of edges, $E$. 
A set $S$ is called a $C$-dismantling set, if the largest connected component of a network 
contains at most $C$ nodes \cite{janson2008, Braunstein2016}. Finding the $C$-dismantling set is a NP-hard problem. 
Current state-of-the-art methods \cite{Morone2015Nature, Braunstein2016, Zdeborova2016, Morone2016, Tian2017NatComm} make the implicit assumption that the cost of node removal is the same for all nodes in a network, regardless of their importance. 
Here thus, we generalize the network dismantling problem in such a way that the cost of removing a node $i$ can be an arbitrary non-negative number $w_i \in \mathcal{R}$ instead of a unit value.
More formally, for a given network $G(V,E)$, we want to find the subset of nodes $S \subseteq  V$ with the minimum cost of removal, which will result in fragmentation into components of size $C$. 
Depending on the system of interest, the cost $w_i$ could represent the amount of energy needed to remove a node, monetary cost of buying or controlling a node, or some other network measure such as the node importance or influence. 
The presented methodology works for arbitrary non-negative weights, but in the absence of other information, here we use the node degree as a proxy for node importance and the associated removal cost.
 Note that, in case of unit costs, the problem becomes equivalent to the standard network dismantling problem \cite{Morone2015Nature, Braunstein2016, Zdeborova2016, Morone2016, Tian2017NatComm}.

\subsection*{Node-weighted spectral cut}
Let us assume that we want to partition the network $G=(V,E)$ in such a way that the nodes from a set $M \subseteq V$ are not connected to the nodes from the complementary set $\overline{M}= V \setminus M$. 
Whether a node $i$ belongs to the set $M$ is represented by the following vector $v \in R^{n}$:
\begin{equation}
v_i:=
\begin{cases}
+1 & \ {i \in M}, \\
-1 & \ {otherwise}.
\end{cases}
\end{equation}
The classical spectral bisection of a graph aims to minimize the number of edges that has to be removed between the clusters $M$ and $\overline{M}$. 
In this paper we propose a novel node--weighted spectral cut objective function, where the cost of cutting the edge $(i,j)$ is equivalent to the cost of removing nodes $i$ and $j$. 
 
Then the upper bound for the cost of removing a subset of nodes that are adjacent to the edges separating clusters $M$ and $\overline{M}$ is:
\begin{equation} 
 \frac{1}{2} \sum_{i,j}  -\frac{1}{2} \left( v_i v_j - 1 \right) A_{i,j} \left( w_i + w_j - 1 \right),
\end{equation}
where the matrix $A$ denotes the adjacency matrix of the network. 
Therefore, if the edge $(i,j)$ connects nodes from different clusters, the associated cost is $w_i + w_j -1$, as $v_i v_j = -1$ and $A_{i,j} = 1$. In contrast, if the edge $(i,j)$ connects nodes from the same cluster ($v_i v_j = 1$), the associated cost is zero, as this link is not removed.
Without loss of generality (see SI section 1 for more details), we assume that the proxy for the weight is proportional to the degree centrality $w_i \propto d_i$.
The term $\left( w_i + w_j -1 \right)$ contains the constant element $-1$
in order to lead to a more elegant notation. Additionally, it corrects for double counting of links that connects $i$ and $j$. 
Now, we define the matrix $B$ by the elements $B_{i,j}=A_{i,j} \left( w_i + w_j -1 \right)$,
and define the node-weighted Laplacian of the matrix $B=AW + WA - A$ by $L_w=D_B-B$. 
In matrix notation the optimization problem can be written as:
\begin{equation}
\label{eq:weg_laplacian}
 \min \frac{1}{4} v^T L_w v
\end{equation}
subject to 
\begin{equation}
\textbf{1}^{T} v = 0,
\end{equation}
\begin{equation}
v_i \in \left\lbrace +1, -1 \right\rbrace, i \in \left\lbrace 1, 2, ..., n \right\rbrace.
\end{equation}
Matrices $W$ and $D_B$ are diagonal matrices
with the elements $W_{ii}= d_i$ and $(D_B)_{ii}=\sum_{j=1}^n B_{ij}$.
For more details about the objective function see section 1 of the SI. 

When the weight matrix equals the identity matrix ($W=I$), we get the unweighted Laplacian, which corresponds to the classical bisection problem \cite{fiedler73, Pothen1990,NewmanSC}. 
The additional constraint $\textbf{1}^{T}v=0$ enforces that clusters are of the same size. Unfortunately, the optimization problem is NP-hard. Therefore, we follow the standard relaxation \cite{fiedler73} from the integer constraint $v_i \in \left\lbrace +1, -1 \right\rbrace$ to $v_i \in \mathcal{R}$. 
The solution to this relaxed constrained minimization problem is, according to the Courant-Fisher theorem, analytically given by the second smallest eigenvector of the node--weighted Laplacian $\lambda_2 v^{(2)} = L_{w} v^{(2)}$. A more detailed derivation of this solution is presented in section 1 of the SI. 
If we remove all the nodes $i$ whose corresponding element in the second smallest eigenvector is non-negative ($v_{i}^{(2)}>=0$) and has a neighbor $j$ with a negative entry ($v_{j}^{(2)}<0$), the network will fragment into two sub-networks $M$ and $\bar{M}$. 
Note that we can make a fine-tuning of the spectral approximation solution, which will be described later.
Recursively, the node--weighted spectral cut is applied to $M$ and $\bar{M}$ until the network is sufficiently fragmented into small subnetworks of maximum size $C$.  



\begin{figure}
\centering
\includegraphics[width=\linewidth]{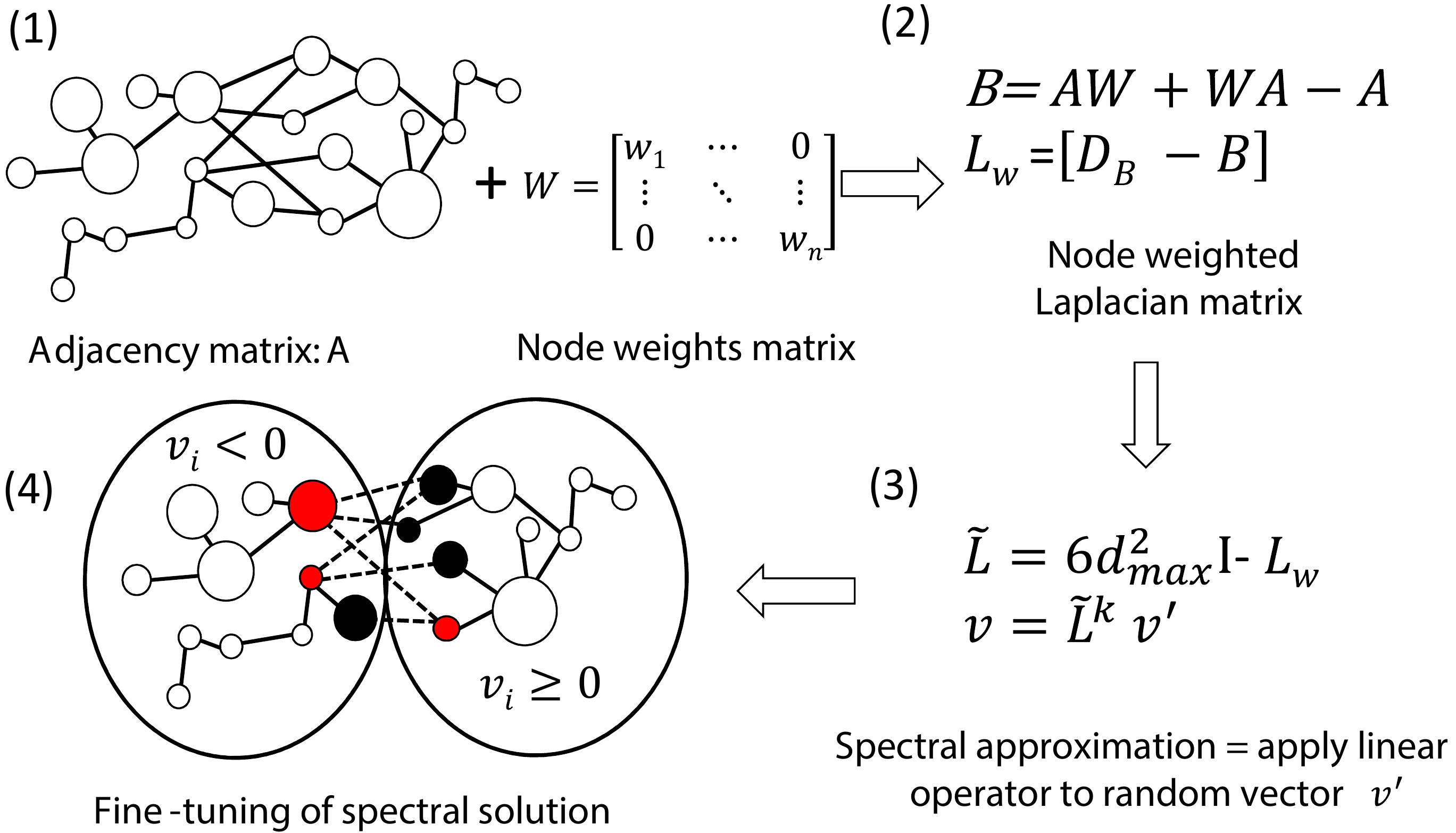}
\caption{Schematic diagram of the proposed GND method for generalized network dismantling. \textbf{(1)} The input network is defined by the adjacency matrix $A$. The costs for node removals are represented by the diagonal matrix $W$ and visualized by different node sizes in the network. \textbf{(2)} Construction of the cost-weighted network defined by the adjacency matrix $B$ and it's corresponding node weighted Laplacian $L_w$. 
\textbf{(3)} Construction of the Power Laplacian operator $\tilde{L}^k$, which is applied to the random vector $v^{\prime}$ on an $n$-dimensional sphere that is perpendicular to the first eigenvector $v_1=(1,1,....,1)$. The result gives an approximate solution to the generalized network dismantling into two components $\left\lbrace i:v_i <0 \right\rbrace$ and $\left\lbrace i:v_i \geq 0 \right\rbrace$. The operator $\tilde{L}^k$ is constructed from the node--weighted Laplacian ($L_W$) and the scaled identity matrix $\textbf{I}$ with the maximal degree of a node in the network, $d_{max}$.
\textbf{(4)} Fine-tuning of the spectral solution is done with the weighted vertex cover on the subgraph of nodes that contains edges between components (represented in black and red). The solution of the fine-tunning is a subset of nodes represented in red.}
\label{fig:shema}
\end{figure}

\begin{figure*}
\centering
\includegraphics[width=1\linewidth]{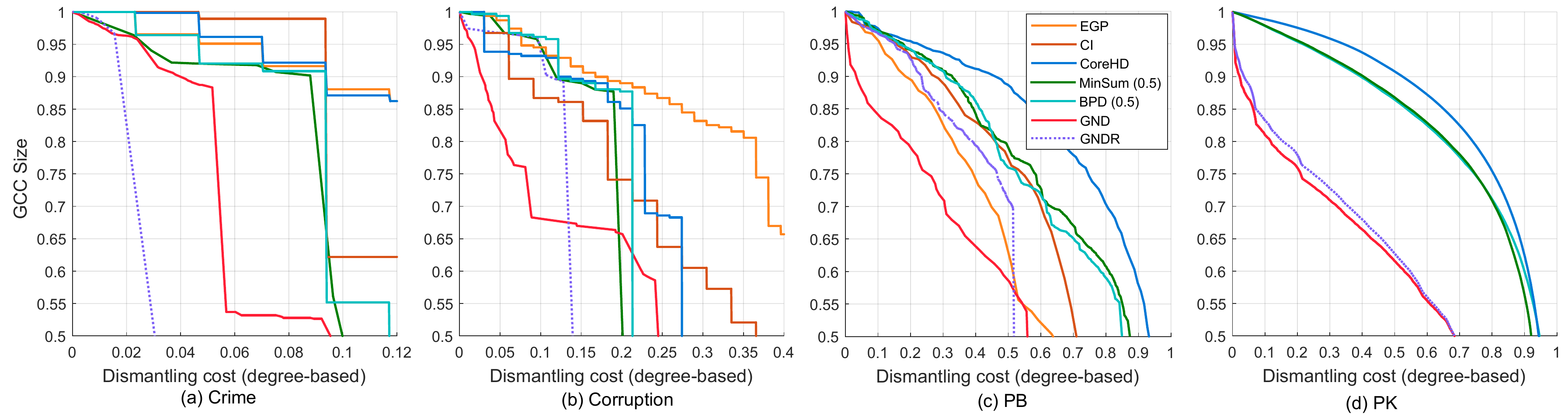}
\caption{ Dismantling of criminal and corruption networks, creation of firewalls to stop the spread of misinformation or malicious software or viruses in online networks. Specifically, we show the size of the GCC versus the overall dismantling cost for four different networks: (a) crime network \cite{konect_crime}, (b) corruption network \cite{Ribeiro2018}, (c) information network (Political Blogs - PB) \cite{Adamic2005} and (d) online social network (Pokec - PK) \cite{takac2012data}. 
In this example, the cost of removing a node is assumed to be proportional to its current degree. The dismantling cost is measured as the fraction of removed edges adjacent to the removed nodes. The network fragmentation was measured for a fixed realistic size from 0.5 to 1.0 (partial dismantling). The results show that our GND and GNDR strategies significantly outperform current state-of-the-art strategies. On the Pokec network with $1.63*10^6$ nodes and $2.23*10^7$ edges, we only compare our methods with Min-Sum, BPD and CoreHD methods, due to the scalability issues of other methods.
}
\label{fig:3Networks}
\end{figure*}

\subsection*{Spectral approximation} 
In order to find the second smallest eigenvectors for large-scale networks, we propose the following simple and elegant approximation algorithm. 
Note that the $L_w$ is a real, symmetric and positive semidefinite matrix. Then, it has real non-negative eigenvalues $\lambda_1 \leq \lambda_2\leq...\leq \lambda_n$ with the eigenvectors $v^{(1)},...,v^{(n)}$, which form an orthonormal basis of $\mathbb{R}^n$. 
In section 2 of the SI, we show that $0= \lambda_1$ and $\lambda_n\leq 6\cdot d_{max}^2$, where $d_{max}$ is the maximum degree of any node of the network. Furthermore, in section 2 of the SI, we also give spectral bound for general non-negative weights $\lambda_n\leq 4 d_{max}(w_{max}+1)$, where $w_{max}$ is the maximum cost.
So, in order to compute $v^{(2)}$, we consider the matrix $\tilde{L}=6\cdot d_{max}^2\cdot I-L_w$, which has the same eigenvectors $v^{(1)},...,v^{(n)}$ as $L_w$. Now the corresponding eigenvalues are shifted such that $\tilde{\lambda_1}=6\cdot d_{max}^2\geq...\geq\tilde{\lambda_n}=6\cdot d_{max}^2-\lambda_n\geq 0$. Let $v^{(1)}$ correspond to the eigenvector with the largest eigenvalue and $v^{(2)}$ to the eigenvector with the second largest eigenvalue. 
Then, we find the eigenvector of $L_w$ associated with the eigenvalue $\lambda_2$ via the following steps: (i) start with a random vector $v$ uniformly drawn from the unit sphere $S^n$, 
(ii) force it to be perpendicular to the first eigenvector $v_1=(1,...,1)^T$ of the weighted Laplacian $L_w$ and (iii) apply the linear operator $\tilde{L}^k$ with unit normalization to our vector $v$. 
The pseudo-code of this spectral approximation is:
\begin{enumerate}
\item Draw $v$ randomly from a uniform distribution on the unit sphere.
\item Set $v = v-\frac{v_1^Tv}{v_1^Tv_1}\cdot v_1$.
\item For $i=1$ to $k = \eta (n)$, set { \\
$v = \frac{\tilde{L} v}{\Vert \tilde{L} v\Vert } $.
}
\end{enumerate}

The intuition that the random vector $v$ converges exponentially to some eigenvector of $L_w$ with eigenvalue $\lambda_2$ is closely related to the spectral properties of operator $\tilde{L}^k$. Note that we can represent our random vector $v$ in the 
orthonormal eigenvector basis as $v = \sum_{i=1}^n \psi_i v^{(i)}$. The second step of orthogonalization ensures $\psi_1 = 0$ and $\psi_2\neq 0$ (almost surely). 
Finally, by applying the linear operator $\tilde{L}^k$ to vector $v$ we get:
\begin{equation}
\label{eq:powerIteration-opertor}
\tilde{L}^k v = \sum_{i=2}^n \psi_i \tilde{\lambda_i}^k v^{(i)} \propto \psi_2 v^{(2)} + \sum_{i=3}^n \psi_i \left(\frac{\tilde{\lambda_i}}{\tilde{\lambda_2}}\right) ^k v^{(i)}.
\end{equation}
When $\lambda_3>\lambda_2$ we have $\vert \frac{\tilde{\lambda_i}}{\tilde{\lambda_2}}\vert<1$, $\left(\frac{\tilde{\lambda_i}}{\tilde{\lambda_2}}\right) ^k\psi_i v_i\rightarrow 0$ with exponential speed. 
The expected value of vector $v$ converges to some eigenvector of $L_w$ with eigenvalue $\lambda_2$:
\begin{equation}
\mathbb{E}\left[\vert \lambda_2 -\frac{v^TL_wv}{v^Tv}\vert \right  ]\to 0,
\end{equation}
when the power $k$ of operator $\tilde{L}$ scales as $\mathcal{O}(\log(n)^{1+\epsilon})$ for every real number $\epsilon >0$,
where $n$ is the size of the network.

If $\lambda_2=\lambda_3=...=\lambda_k <\lambda_{k+1}$, this sequence converges to a unit length linear combination of $v_2,...,v_k$, and is therefore a vector which still minimizes $\frac{v^TL_wv}{v^Tv}$ among all vectors, that are orthogonal to $v_1$. 
 Formal proofs for the convergence and bounds are given in section 3 of the SI.
 
The computational complexity of recursively applying this procedure to smaller and smaller partitions is $O( n \cdot \eta (n)\cdot \log(n))$ for sparse networks.
Due to the fast convergence, one can expect asymptotically good partitions when $\eta (n) = \log(n)^{1+\epsilon}$ and $\epsilon >0$, which finally ends in the complexity of $O( n \cdot \log^{2+\epsilon}(n))$ for sparse networks.
Further details about the asymptotic complexity are given in section 4 of the SI. 

\subsection*{Fine-tuning of the spectral solution}
Let us denote with $E^*$ the set of separating edges that connect nodes from the set $\left\lbrace v_i \geq 0 \right\rbrace$ to the set $\left\lbrace v_i < 0 \right\rbrace$. The set of nodes that are adjacent to the separating set $E^*$ is denoted by $V^*$.
We can optimize the solution by finding a set of nodes which covers all the edges in $E^*$ with minimal cost. This is the {weighted vertex cover problem \cite{BARYEHUDA1981} on the graph $G^*=(V^*,E^*)$ with weights $w_i = \sum_j A_{i,j}$, according to the degrees in the original network $G=(V,E)$. 
This subproblem is also an NP-hard problem. However there exists a 2-approximation efficient solution \cite{BARYEHUDA1981} for it.
Therefore, the cost of the approximate solution for the subproblem is, at most, two times the cost of the optimal fine-tuning. 
Further details about the fine-tuning approximation are provided in section 5 of the SI. A general overview of our proposed method is given in Fig. \ref{fig:shema}, to which we refer as the GND method in the rest of the paper.
At last, as the proposed GND method is offering a recursive solution, some of the nodes from early stages of fragmentation do not contribute to the final stage of complete fragmentation. 
Therefore, in order to produce better dismantling solutions (GNDR) for the complete fragmentation, we reduce some of the nodes from the final dismantling set. More details are provided in section 6 of the SI.

\section*{Results}
In order to demonstrate the applicability of the proposed generalized network dismantling framework to a realistic scenario, we apply it to the real-world networks and show that the current state-of-the-art dismantling strategy \cite{Zdeborova2016} delivers different results from the non-unit cost definition, as expected.
We make the following two realistic assumptions: (i) the cost of removing a node is not constant, but proportional to the importance of the node, measured here by its current degree, and (ii) we focus on the partial dismantling of the system's giant connected component (GCC). 
The first assumption that the cost of removing a node is non-unit 
was already motivated in this paper before. The second assumption reflects the fact that, in practical applications, a partial dismantling of the system size to say, 80 \%, 50 \%, or 1 \% of the original GCC size is more realistic than the complete dismantling, as the budget is usually limited such that only a partial dismantling is possible. 
The degree-based cost of the dismantling is measured by the number of removed edges adjacent to the removed nodes, which is normalized with the total number of edges in the network. In the case of unit-costs, the cost of the dismantling is the fraction of the removed nodes as compared to the total number of nodes in the network.  

In  Fig \ref{fig:motivation}, we show the results of the network dismantling, which represent suppressing the spread of misinformation, computer viruses or other harmful contagion process on the online social network (Petster-hamster \cite{KONECT}). The cost for the 80 \% partial dismantling with the state-of-the-art Min-Sum strategy \cite{Zdeborova2016} is 0.4. 
However,
although the Min-Sum algorithm removes only 5 \% of nodes in this process, its cost is rather large. The reason for this high cost becomes clear if we study the degree distribution of the removed nodes in Fig. \ref{fig:motivation}b, where we notice that all large hubs are removed.
In contrast, the random removal of nodes, also known as site percolation process, with the same cost of 0.4 achieves fragmentation to approximately 75 \% of the original GCC size. 
This implies that the current state-of-the-art strategy becomes very inefficient when the non-unit cost is taken into consideration.
Finally, with the same cost of 0.4, our GND method fragments the network to 62 \% of the original GCC size, and for the target of 80 \% of the GCC size, the corresponding cost is only 0.2.

Next, we study the partial dismantling up to 50 \% of the GCC size on four different real-world networks for five different state-of-the-art methods: Equal Graph Partitioning (EGP) \cite{Chen2008EGP}, Collective Influence (CI) \cite{Morone2015Nature}, Min-Sum \cite{Braunstein2016}, CoreHD \cite{Zdeborova2016} and Belief Propagation-guided Decimation (BPD) \cite{BPAttack}. 
The real-world networks include: (i) crime network with 754 nodes obtained by the projection of a bipartite network of persons and crimes \cite{konect_crime}; (ii)  corruption network \cite{Ribeiro2018} with 309 nodes; (iii) information network of political blogs (PB) \cite{Adamic2005} with 1222 nodes and 16714 edges; and (iv) large online social network (Pokec) \cite{takac2012data} with $1.63*10^6$ nodes and $2.23*10^7$ edges.
In a case of malfunction, the dismantling of these networks can enable efficient immunization strategies against harmful contagion by engineered breaking points in criminal or corruption networks, or by
firewalls to stop the spread of misinformation and malicious cyber data.
In Fig. \ref{fig:3Networks}, the results show that, for the partial dismantling to 50 \% of the original GCC size, the proposed methodology (GND and GNDR) achieves the same fragmentation level with much smaller cost: 
$0.03$ (GNDR) vs. $0.1$ (Min-Sum) for the crime network, 
$0.14$ (GNDR) vs. $0.19$ (Min-Sum) for the corruption network, 
$0.55$ (GND) vs. $0.65$ (EGP) for the information network and 
$0.69$ (GND) vs. $0.91$ (Min-Sum) for the online network. 
If we did the unit cost dismantling analysis, our approach was still better or comparable to the other approaches (see section 7 of the SI for more details). 
In Fig. \ref{fig:petsterFull}, we show the fragmentation curve for the complete dismantling (approximately 1 \% of the original GCC size) for different weighting: (a) degree-based cost, and (b) unit costs.
For unit costs, our approaches provide better or comparable solutions.

\begin{figure}
\centering
\includegraphics[width=\linewidth]{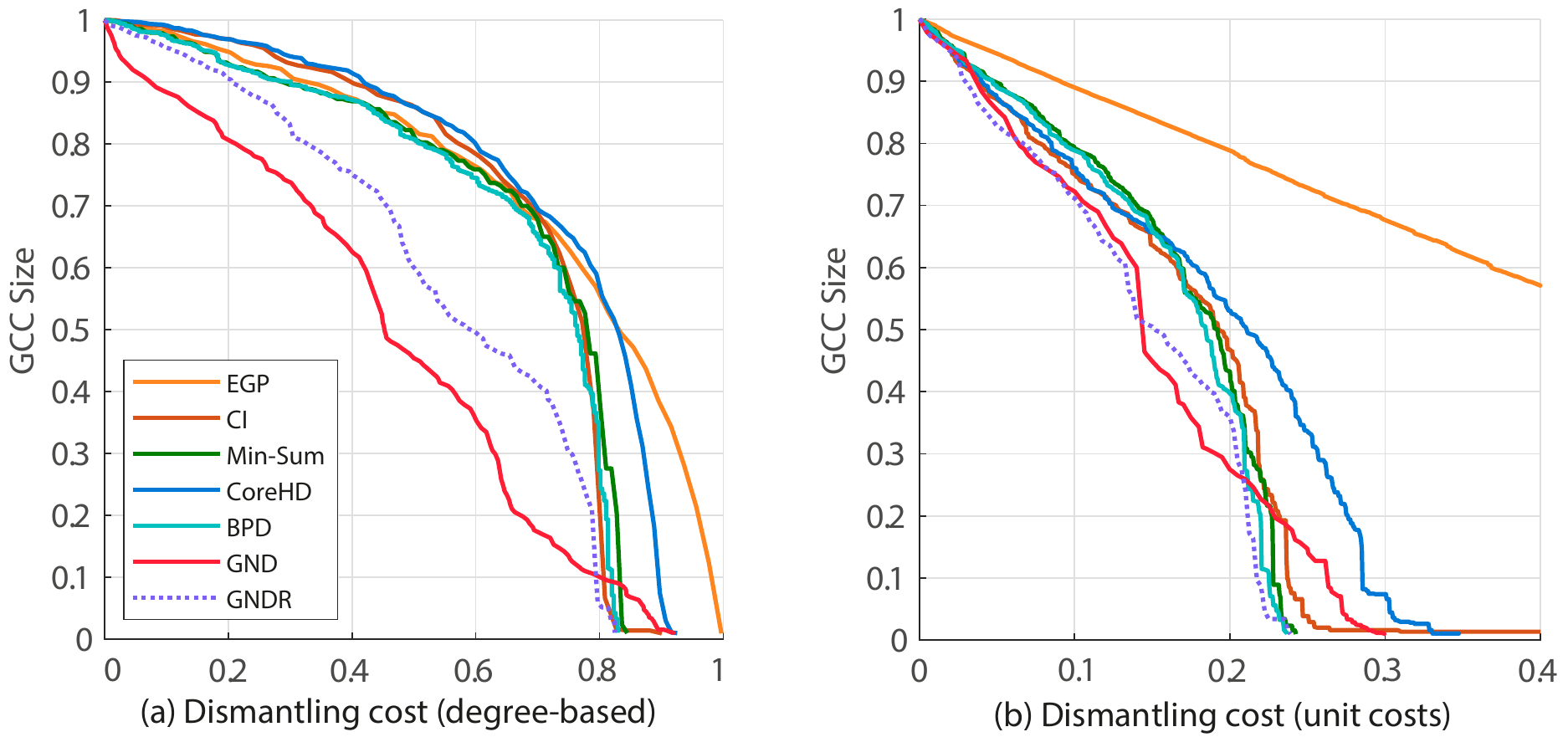}
\caption{Dismantling represents the controlled process of suppressing the spread of misinformation, computer viruses or other harmful contagion processes on the online social network (Petster-hamster \cite{KONECT}). Size of the GCC versus the dismantling cost for complete dismantling (target size $0.01$). 
The cost of removing a node is: (a) proportional to the current degree of a node or (b) equal for all nodes (unit costs). We observe that even for unit costs and complete dismantling the presented methodology (GNDR) provides good solutions.}
\label{fig:petsterFull}
\end{figure}


\begin{figure*}
\centering
\includegraphics[width=1\linewidth]{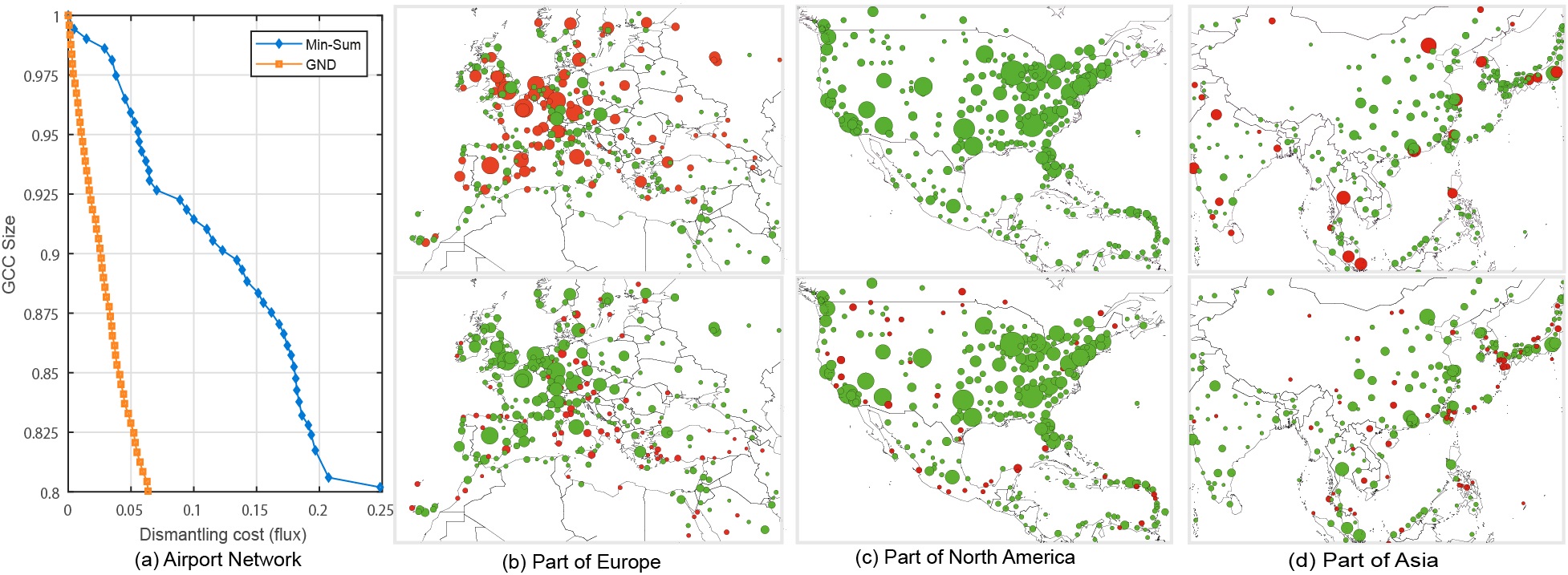}
\caption{Comparison of our proposed GND algorithm with the Min-Sum algorithm for the world airport network. The cost of a node $w_i$ in this network is assumed to be given by the total passenger flux of the airport. (a) Setting the target size of the GCC of the network to 80\%, the Min-Sum algorithm \cite{Braunstein2016} implies a cost of approximately 25 \% of the total passengers. In contrast to the Min-Sum algorithm, our GND method dismantles the network with the cost of only 6 \% of the total passengers, which 
amounting to 19 \% less for the same target dismantling size. 
(b), (c), and (d) visualize the airports that will be closed for a target size of 80\% of the GCC size using the Min-Sum algorithm (upper panel) or the GND algorithm (lower panel) in Europe, North America, and Asia, respectively. Closed airports are represented by red dots with the area proportional to the cost.}
\label{fig:airport}
\end{figure*}

In case the external information about the cost of removing a node $w_i$ is available, we are able to incorporate it into the matrix $W$ and proceed with our GND method. Section 2 of the SI gives spectral bounds for general non-negative weights, for which the same spectral approximation method can be used.
In Fig. \ref{fig:airport}, we 
show the results for the world airport network, where the cost $w_i$ of closing an airport $i$ is assumed to be proportional to the total flux of the passengers of the airport. In this example, we have set the target size of the GCC of the network at 80\% of the initial size. It is interesting to observe that our GND method dismantles the network with the cost of only 0.06 of the total weights, which is significantly less than the cost of 0.25 incurred with the Min-Sum method \cite{Braunstein2016}.
In the same figure, we provide a geographical visualization of the dismantling solution, where the closed airports are represented by red dots with area  proportional to the dismantling cost.  
The world airport case study shows the importance of considering realistic dismantling costs, which dramatically changes the dismantling solutions. In case of the world airport network, the closing of an airport can represent quarantine. Correspondingly, the reduction of the GCC size represents the containment effect for the pandemic spread.

\section*{Conclusion}
In this paper, we introduce the generalized network dismantling problem, which seeks to find a set of nodes allowing to dismantle a network into
components of subcritical size in the most cost-effective way. 
We do not make the assumption that the cost of removing nodes is the same for all nodes \cite{Morone2015Nature, Kovcs2015, Braunstein2016, Zdeborova2016, BPAttack, Morone2016, Tian2017NatComm, Zhou2013FVS}. 
We allow for costs to include non-topological properties related to the price or protection level.
Our proposed method is based on a blend of spectral properties of a novel node-weighted Laplacian operator, randomized approximations and weighted vertex cover approximations. 
We demonstrate that, for the partial dismantling of networked systems, current state-of-the-art methods do not produce near-optimal results and sometimes behave even worse than the random baseline strategy (site percolation method). 
Our study raises new questions regarding the reorganization of current socio-technical systems under different realistic costs in order to become more robust against targeted attacks.
We have demonstrated that the dismantling can enable cost-effective immunization strategies against harmful contagion effects in social and transportation networks as well as the disruption of criminal and corruption networks. 
Understanding the theory behind network dismantling opens new research directions and will enable us to design more robust and resilient systems in future.

\subsection*{Ethics}
The method presented in this paper aims at offering a possible solution for emergencies where cutting a dysfunctional network into pieces can restore the functionality. However, we also warn of potential misuses or dual uses. When not applied in appropriate contexts and ways, the use of the dismantling approach may undermine the proper functionality of networks. Therefore, we point out that related ethical issues must be always sufficiently, appropriately, and transparently addressed \cite{helbing2017will, EthicsDataScience} when the method is applied. The method must be restricted to legitimate uses and actors. It may be justified to stop harmful cascading problems such as deadly epidemics and the spreading of disruptive computer malware, or to dismantle criminal organizations or corruption networks. The method may also be used to identify more resilient system designs and network operations. Note, however, that the use of dismantling strategies to contain misinformation can be potentially problematic, as it may result in censorship if a government, company, news agency or other institution decides what is misinformation or not. Stopping the spread of true information can seriously obstruct the societal evolution towards better insights and solutions. 
(For example, if the method had been misused by the established powers in the past, we might still believe that the Earth is the center of the universe.)
Also note that, if public discourse is shaped by a few people only, this may promote the misuse of power, corruption and crime. In order to contain fake news, dis- and misinformation, we recommend a suitable combination of the use of AI, collective intelligence (such as Wikipedia and crowd-sourced fact checking), reputation systems for messages and information sources, elected community moderators, complaint mechanisms, qualification mechanisms, quality-based message ranking and reach, as well as verification/measurement-based approaches.


\acknow{X.-L.R. thanks for financial support by China Scholarship Council (CSC). N.A.-F. and D.H. are grateful for financial support from the EU Horizon 2020 projects: SoBigData under grant agreement No. 654024 and CIMPLEX under grant agreement No. 641191. Authors would like to thank to A. Lancic and K.K. Kleinberg for useful comments and suggestions. 
}

\showacknow{} 


\bibliography{pnas-sample}

\end{document}